\documentclass[12pt]{iopart}

\usepackage{etoolbox}
\usepackage[utf8]{inputenc}

\makeatletter
\newcommand{\mainmatter}{%
  \setcounter{footnote}{0}%
  \patchcmd{\@makefntext}{\fnsymbol}{\arabic}{}{}%
  \patchcmd{\@thefnmark}{\fnsymbol}{\arabic}{}{}%
  \def\@makefnmark{\textsuperscript{\arabic{footnote}}}%
}
\makeatother

\usepackage{iopams}

\expandafter\let\csname equation*\endcsname\relax

\expandafter\let\csname endequation*\endcsname\relax

\usepackage{mathrsfs,mathtools,bbold}
\usepackage{amsmath,amssymb}
\usepackage{amsfonts}
\usepackage{float, cite}
\usepackage[colorlinks,linktocpage]{hyperref}
\usepackage[dvipsnames]{xcolor}
\usepackage{physics}
\usepackage{slashed}
\usepackage{bm, cancel}
\usepackage[singlelinecheck=false 
]{caption}
\usepackage[flushleft]{threeparttable}
\usepackage{multirow}
\usepackage{booktabs}
\usepackage{verbatim}

\hypersetup{colorlinks=true,linkcolor=blue,citecolor=blue,urlcolor=blue}

\newcommand{\sqrtP}[1]{\sqrt{\smash[b]{#1}}}

\begin{document}
\title[Pseudoscalar corrections to spin motion equation]{Pseudoscalar corrections to spin motion equation, search for electric dipole moment and muon magnetic (g-2) factor}

\author{V. G. Baryshevsky$^1$ and P. I. Porshnev$^2$}
\address{$^1$ Institute for Nuclear Problems, Belarusian State University, Minsk, Belarus}
\address{$^2$ Past affiliation: Physics Department, Belarusian State University, Minsk, Belarus}
\vspace{10pt}
\begin{indented}
\item[]November 2020
\end{indented}

\begin{abstract}
The spin dynamics in constant electromagnetic fields is described by the Bargmann-Michel-Telegdi equation which can be upgraded with anomalous magnetic and electric dipole moments. The upgraded equation remains self-consistent, Lorentz-covariant and gauge-invariant. It and its different forms have been confirmed in numerous experiments to high degree of accuracy. We have recently derived the spin motion equation within the Wentzel–Kramers–Brillouin weak-field approximation which adds a pseudoscalar correction to the BMT equation. The upgraded equation is again self-consistent, Lorentz-covariant, gauge-invariant, and free of  unwanted artifacts. The pseudoscalar correction is expected to be small, and might become important in hypersensitive experiments, like the measurements of electric dipole moments which are themselves related to pseudoscalar quantities. It also becomes possible to explain why EDMs are so difficult to measure, since this correction term might lead to the effective screening of electric dipole moments. Within the same model, it is possible to explain the discrepancy between experimental and theoretical values of muon magnetic anomaly under assumption that the pseudoscalar correction is the dominant source of this discrepancy.
\end{abstract}

\vspace{2pc}
\vspace{10pc}
{\footnotesize\color{gray} \ead{ \,$^1 \text{v\_baryshevsky@yahoo.com}$\\$\qquad\qquad^2 \text{pporshnev@gmail.com}$}}
%
%

\maketitle

\date{July 15, 2020}

\section{Introduction}
The motivation behind this work could be explained as follows. The search for electric dipole moment (EDM) has been going for several decades. The outstanding progress has been made in reducing the upper bounds by many orders of magnitudes; still, no EDMs have been detected so far. The ongoing search has two remarkable sides to it. From the one hand, field theoretical extensions to standard model show a good breadth in predictions and possible mechanisms of CP-violations, many of which have been already rejected. If a nonzero EDM is discovered one day, there is a good chance that one of already existing extensions could be used to describe it. From the other hand, the phenomenological models that are used for interpreting the experimental data have been quite rigid. Their essential features were established many decades ago.

The Bargmann-Michel-Telegdi (BMT) equation and non-second-quantized Hamiltonians  which are both upgraded with anomalous magnetic (AMM) and electric dipole moments are two commonly used phenomenological models.\hspace{-7pt}
\footnote[7]{\color{gray} Depending on experimental set-ups, additional terms that include interactions with other particles, nuclei, or lattice atoms can be included.  However, these additional terms, even if some of them are $CP$-violating, are not directly related to the EDM problem of individual particles. We also consider the weak-field limit, and ignore terms with field gradients. It is not what we target in this study. See more discussion below.}
The validity of these models is based on the solid foundation: there is a great deal of experimental data, including atomic data and very precise measurements of AMM, to support them. Still, from purely a logical ground, venturing into the area of new hypersensitive measurements, there is no guarantee that some modification of established phenomenological models would be required, since we might be leaving their support area.  

Is there an opportunity to extend the phenomenological description in a consistent way without running into problems with existing experimental data? The requirements are very strict. Consider, for example, the motion and BMT equations with AMM and EDM terms which are Lorentz-covariant, gauge-invariant, and free of artifacts, like zitterbewegung or runaway acceleration.  An  extended model should not lose these properties while adding potentially new corrections to areas that have not been thoroughly tested before. Additionally, any corrections to well established quantities, like anomalous magnetic moment, must be below or at the currently tested uncertainty level.  Remarkably, such an extension to existing spin motion equations can still be constructed. The extended model that is proposed \cite{baryshevsky_search_2020, porshnev_electron_2020} is self-consistent, Lorentz-covariant, gauge-invariant and free of unwanted artifacts. 

Inspecting Lagrangian terms in effective QFTs, we see that all sixteen gauge-invariant Dirac bilinears (products of fields) are used, either as standalone or in coupling with other particle currents, see table \ref{eedm_table}. Accordingly, the commonly used phenomenological models include most of corresponding terms or coefficients. The BMT equation 
\begin{align}\label{eedm_21}
	\dv{s^\mu}{\tau} 
	=\frac{ge}{2m} F^{\mu\nu}s_\nu+\frac{a e}{m}( s^\rho F_{\rho\nu}u^\nu)  u^\mu\,,
\end{align}
can be viewed as defined by means of nine Dirac bilinears or two normalized four-vectors
\begin{align}
	&u^\mu = \frac{1}{\bar{\psi}\psi}\bar{\psi}\gamma^\mu\psi\,, &s^\mu =\frac{1}{\bar{\psi}\psi}\bar{\psi}\gamma^\mu\gamma^5\psi\,,	
\end{align}
where $u^\mu$ is the four-velocity, $s^\mu$ is the expectation value of Dirac pseudovector $\bar\psi\gamma^\mu\gamma^5 \psi$, $F^{\mu\nu}$ is the electromagnetic tensor, $a=g/2-1$ is the anomalous magnetic moment, and $\bar{\psi}\psi$ is the density. The BMT equation  includes the $g$-factor which origin is traced to the bilinear tensor $\bar\psi\sigma^{\mu\nu} \psi$. Correspondingly, the term with AMM can be included into the one-particle Hamiltonian together with other corrections, including the relativistic ones. 
\begin{table}[h!]
\centering
  \begin{threeparttable}
\caption{Lagrangian terms in effective QFT theories and corresponding terms in phenomenological models of spin motion.}    
\begin{tabular}{lccccccccccccccc}
\hline
\hline
					&   	&&  \multicolumn{3}{c}{Phenomenological models}\\ \cmidrule{4-6}
Bilinear			& QFT	&&	$\text{BMT}$			& $\text{BMT}+ d$	& \cite{baryshevsky_search_2020} \\
\hline
$\bar\psi \psi$							&$\ast$	&&$\ast$	&$\ast$	&$\ast$\\
$\bar\psi\gamma^\mu \psi$    			&$\ast$	&&$\ast$	&$\ast$ &$\ast$\\
$\bar\psi\gamma^\mu\gamma^5 \psi$		&$\ast$	&&$\ast$	&$\ast$	&$\ast$\\
$\bar\psi\sigma^{\mu\nu} \psi$			&$\ast$	&&$\ast$	&$\ast$	&$\ast$\\
$\bar\psi\gamma^5\sigma^{\mu\nu} \psi$	&$\ast$	&&	 		&$\ast$	&$\ast$\\
$\bar\psi i\gamma^5 \psi$				&$\ast$	&&			&  		&$\ast$\\
\hline

\end{tabular}
\label{eedm_table}
%
  \end{threeparttable}
\end{table}

Analogously, if an effective Lagrangian includes the $CP$-odd term with electric dipole moment $d$
\begin{equation}
	L_{eff} = \dots + id F_{\mu\nu}\bar\psi\sigma^{\mu\nu}\gamma^5\psi\qquad\to\qquad \dv{s^\mu}{\tau}=\dots- 2d\qty( \tilde F^{\mu\nu}  s_\nu+  s_\rho \tilde F^{\rho\nu}  u_\nu u^\mu )\,,
\end{equation}
where the dual field  tensor $\tilde F^{\mu\nu}$ is defined as  $\varepsilon^{\mu\nu\rho\sigma}F_{\rho\sigma}/2$,
the corresponding terms are added to the phenomenological description of spin motion. Hence, the contribution of another set of Dirac bilinears  that are given by the tensor $\bar\psi\gamma^5\sigma^{\mu\nu} \psi$ in the Lagrangian is captured with the coefficient $d$ in the phenomenological model.

The pseudoscalar bilinear $q=\bar\psi i\gamma^5 \psi$ is explicitly absent in commonly used phenomenological models of spin dynamics, even if terms with pseudoscalar currents are explicitly used in effective Lagrangians of  various extensions to the standard model. The clarification is in order. The pseudoscalar $q$ is a gauge-invariant quantity that belongs to the same class of well-defined physical quantities as the fermion density, current, or spin four-vector.  A change in fermion motion under influence of external fields causes corresponding changes in all Dirac bilinears, including the pseudoscalar. It can be called the dynamically induced pseudoscalar, since free fermions  are assumed to have zero pseudoscalar in conventional models. Let us write it this way
\begin{equation}
	q = q_0 + q_{dyn}(\delta \psi)\,,
\end{equation}
where $q_0$ is the static pseudoscalar, and $q_{dyn}$ is its dynamic part. The argument of $q_{dyn}$ is meant to show that it is induced by changes in fermion wave function, excluding changes in its phase. For a field-free case and constant $\psi$, the dynamic part turns zero. We conventionally assume that for free fermions the pseudoscalar is zero; saying otherwise the static part $q_0$ is strictly zero.  This assumption is typically made implicitly. 

A fermion is described by several fundamental constants
\begin{equation}
	(e, m, s, a, d, q_0, \dots)
\end{equation}
which are charge, mass, spin, magnetic and electric moments. The rest  includes the weak isospin, decay times, and other relevant properties. We add $q_0$ to this list for the following reason. The pseudoscalar is well defined physical quantity which is real-valued, gauge-invariant, and Lorentz invariant as charge and mass. Even if $q_0$ is strictly zero, it is still the statement about the properties of free fermion state. Its zero value is still the assumption which is a subject to experimental verification. We will discuss below that it is the very good assumption in fact; it has been supported by available data so far. Our main hypothesis is that $q_0$ is very small but it might differ from zero. More, the phenomenological model that we developed \cite{porshnev_electron_2020, baryshevsky_search_2020} allows to experimentally test this hypothesis and opens new opportunities in measurements of both EDM and AMM coefficients.

From the field point of view, the fermion condensate is typically assumed to be the CP-invariant condensate $\langle \bar\psi\psi\rangle$.  What we argue is that some minor fraction of fermion condensate can be a $CP$-noninvariant. Saying otherwise, a real fermion condensate has a small $CP$-noninvariant part  $\langle \bar\psi i \gamma^5\psi\rangle$. Since the pseudoscalar in theories with massive fermions is the source of axial current divergence, the chiral anomaly that is also the source of axial current non-conservation must be directly related. An interaction with a background pseudoscalar field (axions) \cite{peccei_strong_2008,barth_cast_2013} could also be a mechanism for generating such a small symmetry-violating fraction. A detailed discussion of specific mechanisms behind a nonzero static pseudoscalar is outside the scope of this work. Our approach here is strictly phenomenological. We managed to find the way to extend the phenomenological description while keeping all the benign features of conventional models. This extension now includes the parameter $\beta$ which is the ratio of fermion pseudoscalar density to its regular density. Both densities are the same type of bilinears as expectation values for spin and momentum, and their inclusion into the spin equation keeps it covariant, gauge-invariant and free of unwanted artifacts.

The dynamically induced part $q_{dyn}$ is the key part of standard Dirac formalism; we will briefly review it in one of the sections below. It is not the pseudoscalar correction that we are talking about in this study.  Only taking into account that the static pseudoscalar might be nonzero,  small corrections to the EDM and AMM coefficients, in the form of $a\beta$ and $d\beta$ respectively, appear in the spin motion equation \cite{porshnev_electron_2020, baryshevsky_search_2020}. It is important to emphasize that they  appear in the spin motion equation  if we include the anomalous magnetic and electric moments, which themselves are originated by radiative corrections, into the \emph{squared} Dirac equation. We will review it in detail below. There is the clear practical argument why these corrections are expected to be very small.

The well-known spin-dependent Hamiltonian part \cite{berestetskii_quantum_1982}
\begin{equation}\label{eq27}
	H = - \frac{e g}{2m}\hat s\textbf{B}+ \frac{e(g-1)}{2m}\hat s(\textbf{v}\cross\textbf{E})\,,
\end{equation}
where $\hat s=\bm{\sigma}/2$ is the spin operator, has been confirmed experimentally with high accuracy in innumerous atomic experiments \cite{khriplovich_equations_1999}. It does not include either electric dipole moment or pseudoscalar correction. Hence, a hypothetical contribution from $ a\beta$ can only be  a small fraction of radiative corrections that are not captured in the Hamiltonian \eqref{eq27}. Since the anomalous magnetic moment of electron itself is confirmed down to $10^{-12}$ level \cite{aoyama_theory_2019}, any new correction could not exceed these well established bounds. Consequently, the electron pseudoscalar correction can be safely ignored for the most of atomic physics. It could be important only for hypersensitive experiments, like the EDM ones, or for experiments with heavier fermions.

Spin motion  equations can be derived or justified in many different ways: by using the heuristic arguments \cite{berestetskii_quantum_1982, khriplovich_feasibility_1998, fukuyama_derivation_2013,metodiev_thomas-bmt_2015}, with the Foldy–Wouthuysen (FW) transformation \cite{silenko_quantum-mechanical_2005} or within the Wentzel–Kramers–Brillouin (WKB) approximation to the Dirac equation \cite{rubinow_asymptotic_1963,rafanelli_classical_1964}. For homogeneous weak fields, the consistent application of these methods leads to the BMT equation which can be upgraded with both anomalous magnetic and electric dipole moments. A large literature has been dedicated to this topic since the BMT equation was published 60+years ago \cite{bargmann_precession_1959}; the topic itself goes back to the Frenkel's work \cite{frenkel_elektrodynamik_1926}. Most of the publications belong to the heuristic and FW-transformed  Hamiltonian-based derivations. The latter approach is often viewed as the consistent way to obtain spin equations that avoids potential pitfalls that might be encountered outside the Hamiltonian-based approach \cite{suttorp_covariant_1969, suttorp_covariant_1970}. It captures well the dynamic contribution of pseudoscalar into the spin dynamics, however it cannot capture the static pseudoscalar correction in principle, per its design. We have recently re-derived the spin motion equation within the weak-field WKB approximation \cite{porshnev_electron_2020, baryshevsky_search_2020} to the squared Dirac equation.  It matches the conventional BMT equation extended with both AMM and EDM terms \cite{ford_report_1961,mane_spin-polarized_2005}, and additionally introduces a small pseudoscalar correction to those moments. 

The developed phenomenological model allows to potentially explain why the EDM has been so difficult to detect, since this correction term might lead to the effective screening of electric dipole moments. Within the same model, it is possible to explain the discrepancy between experimental and theoretical values of muon magnetic anomaly under assumption that the pseudoscalar correction is the dominant source of this discrepancy.

The paper is organized as follows. In the next section, the pseudoscalar role in the Dirac formalism will be briefly reviewed. Next, we review the Dirac formalism with the AMM  and EDM terms within the WKB approximation. The predictions that might follow from the proposed phenomenological model will be discussed in the final section. 

\section{Dirac pseudoscalar}
The unique nature of pseudoscalar bilinear and its role in the Dirac formalism can be illustrated this way. Consider, for example,  a linear superposition of particle  and antiparticle modes
\begin{equation}\label{eq85}
	\psi = A \mqty(\phi\\ 0) + B \mqty(0\\ \chi)\,,
\end{equation}
where we use the standard representation of Dirac gammas; $\phi$ and $\chi$ are undotted and dotted unit spinors respectively. The pseudoscalar estimates to
\begin{equation}
	q = i\bar\psi \gamma^5\psi = i (A^* B \phi^\dagger\chi - AB^* \chi^\dagger\phi )\,.
\end{equation}
If the fermion is in a pure particle $(B=0)$ or pure antiparticle state $(A=0)$, then $q$  is strictly zero. It is the only covariant bilinear quantity (out of five sets of bilinears with distinct Lorentz covariant properties - density, pseudodensity, vector, pseudovector, and tensor) with such a property! All other Lorentz-covariant sets from table \ref{eedm_table} which are gauge-invariant are nonzero for either pure particle or antiparticle states.

It is important to elaborate that we are still in the one-particle representation; the expression \eqref{eq85} is a general state of single fermion. Conventional phenomenological models describe a general state of single free fermion with only fifteen Dirac bilinears: density, momentum, spin, and antisymmetric tensor (which is called the spin tensor in some models) while  assuming that the remaining pseudoscalar bilinear for free states is zero. From this angle, our extension is maximal since it uses all sixteen gauge-invariant bilinears of single fermion to describe its free states. \hspace{-6pt}\footnote[7]{\color{gray}Extensions that use non gauge-invariant bilinears unavoidably lead to unwanted artifacts.}
Any other extension would require multi-particle states or the field approach.

A bispinor is the direct sum of two irreducible representations $(1/2,0)\oplus(0,1/2)$ of the rotation group. Two spinors represent the pure particle and pure antiparticle modes respectively. Each spinor transforms independently of one another under rotations, however they transform via each other under boosts. This is the reason why the direct sum is required. In such a case, bispinors are transformed linearly and homogeneously under Lorentz transformations. The striking feature of this behavior is the following. If we set one spinor to zero, let it be the bottom one $\chi=0$, then both pseudoscalar and three-dimensional part of four-velocity or current turn zero
\begin{equation}\label{eq28}
	k^\mu  = \bar{\psi}\gamma^\mu\psi=(\abs{\phi}^2 + \abs{\chi}^2,\,\, \phi^*\bm{\sigma}\chi + \chi^*\bm{\sigma}\phi) \quad\to \quad (\abs{\phi}^2,\,\, 0,0,0)\,.
\end{equation}
Effectively, by setting one spinor to zero, we bring the corresponding bispinor to its rest frame and force it into one of two pure modes. Since the time component of $k_\mu$ in \eqref{eq28} is not zero, it can acquire the space part in other inertial frames connected by Lorentz transformations. However, the corresponding pseudoscalar will remain zero in such inertial frames. The opposite is not true: there exists an infinite number of conditions for which $q\ne0$ while $\textbf{k}=0$. It means that a general fermion can have a nonzero pseudoscalar in its rest frame.

It follows then almost trivially that any solution of the field-free Dirac equation
\begin{equation}\label{eq29}
	(i\gamma^\mu - m )\psi_+(x) = 0 \qquad \to\qquad  (\slashed p - m)u_p = P_-u_p = 0\,,
\end{equation}
has zero pseudoscalar.  Here $\psi_+(x)$ is a positive energy mode, $P_{\pm}=(\slashed p \pm m )/2$ are two orthogonal energy projectors, and $u_p$ is the Fourier image of $\psi_+(x)$. A general solution in the momentum space is given by projecting out the positive energy mode $u_p = P_+ \psi(p)$ out of arbitrary bispinor $\psi(p)$. The projector $P_+$ always projects out the upper spinor in the rest frame (standard gammas) by setting the lower one to zero. Since such a solution is given by the bispinor $u_p$ which lower spinor is strictly zero in the standard representation in the rest frame, its corresponding pseudoscalar is zero too. All such solutions of \eqref{eq29} will have zero pseudoscalar, even if one constructs a general wave packet out of different $u_p$. 

Imagine now a case where an electron travels through a weak field $A_\mu$ or the field has been adiabatically turn on  without generating pairs. The electron motion is described now by the minimally coupled Dirac equation
\begin{equation}\label{eq30}
	(i\gamma^\mu - m )\psi(x) - e A_\mu(x) \psi(x) = 0\,. 
\end{equation}
The field will force a change in electron motion, and $\psi$ will develop a nonzero lower spinor which in turn will lead to non-zero pseudoscalar which we call the dynamically induced one.  All other bilinears will evolve as well. When the electron enters a free-field zone or the field is turned off, the particle starts moving with a constant momentum which is different from its original one in general. However, even in this new state the electron must have zero pseudoscalar again, since it still must satisfy the equation  \eqref{eq29}. 

Analogously, if we are dealing with antiparticles, their free motion is described as
\begin{equation}\label{eq31}
	(i\gamma^\mu + m )\psi_-(x) = 0 \qquad \to\qquad  (\slashed p + m)v_p = P_+v_p = 0\,,
\end{equation}
which solutions all have zero pseudoscalar. Now however, the solution is given by the second projection  $v_p = P_-\psi(p)$ which eliminates the upper spinor in the rest frame. The rest of arguments follows almost ad verbatim from the previous paragraph assuming the signs are swapped. 

Next, consider a typical  quantum field process where an electron motion  is accompanied by virtual pairs, and which invariant amplitudes include contributions from both fermions and antifermions because of vacuum polarization. Within the QED framework \cite[p. 247]{peskin_introduction_1995}, a bare electron charge $e_0$ is reduced by vacuum polarization to the effective charge $e_{eff}$ 
\begin{equation}
	\frac{e_{eff}^2}{4\pi} = \frac{e_0^2}{4\pi} \frac{1}{1 - \Pi(k^2)- \Pi(0)}\,,
\end{equation}
where $\Pi(q^2)$ is the vacuum polarization function at transferred momentum $k$. It is not physical by itself unless we subtract its divergent part $\Pi(0)$. The finite part describes then the bare charge reduction by virtual electron-positron pairs. This cloud of virtual pairs around an electron has observable consequences which can be estimated by the change in Coulomb potential (Uehling). The corresponding contribution into the relative shift (Lamb) of hydrogen energy levels is very small ${\sim}10^{-7}eV / 10\, eV$ \cite[p. 253]{peskin_introduction_1995}. Assuming that such a shift is caused by the reduction in the electron charge by virtual positrons, the very crude estimate for this specific case is given as $\beta = q/r \approx \abs{A}/\abs{B} \,{\sim}10^{-8}$ where $r$ is the regular fermion density.

Anyway, since  electrons are surrounded by vacuum charges, including their antiparticles, they might possess a nonzero static pseudoscalar even in the absence of external forces. 
It is the main hypothesis of this work.  The above discussion however shows that we face the challenge of how to describe such a situation with a phenomenological model.  Neither \eqref{eq30} or the analogous equation for antiparticles possesses solutions with nonzero static  pseudoscalars. This is the reason why the conventional and well established derivations of spin dynamics based on \eqref{eq30} do not lead to the static pseudoscalar correction. 

The way to develop a phenomenological model of spin equation with desired properties is to use the squared Dirac equation instead \cite{berestetskii_quantum_1982, baier_radiative_1972, feynman_theory_1958, rafanelli_classical_1964, strange_relativistic_1998,bagrov_squaring_2018}
\begin{equation}\label{eq120}
	(i\slashed\partial -e\slashed A +m)(i\slashed\partial -e\slashed A -m)+\underbrace{\dots}_{\text{field terms}}\quad \to\quad    P_-P_+ \psi = 0\,.
\end{equation}
This well-known form properly recovers the term with $g$-factor and is frequently used in various applications \cite{berestetskii_quantum_1982}. Since it also recovers the BMT equation in the classical limit \cite{rubinow_asymptotic_1963,rafanelli_classical_1964} and the Pauli term for the nonrelativistic motion \cite[p. 120]{berestetskii_quantum_1982}, it can be seen as equally confirmed by atomic phenomenology as the original Dirac equation. In the context of this study however, its main benefit is its  field-free solutions can now have a nonzero pseudoscalar
\begin{equation}
	P_-P_+ \psi= P_-P_+\qty[\mqty(\phi\\0)+\mqty(0\\\chi)] = P_-P_+( P_+ \psi + P_-\psi) = 0\,,
\end{equation}
since it allows both nonzero spinors in the free case.  It is important to emphasize the following. Conventionally, the squared Dirac equation \eqref{eq120} is used as an alternative to the regular one \eqref{eq30} even if the former equation has more solutions compared to the latter one. To ensure the mathematical equivalency, only solutions of second-order equation that also satisfy the first-order equation  \eqref{eq30} are conventionally chosen \cite[p. 120]{berestetskii_quantum_1982}. However, in doing so, we will end up with exactly the same situation as for the regular Dirac equation. The solutions of squared equation that are filtered in such a way will have zero static pseudoscalar; we do not gain anything new then.  Therefore, we will not impose this additional requirement on the solutions of squared Dirac equation; it admits new types of solutions compared to the regular approach.
It should be mentioned that such new solutions can now have nonzero pseudoscalar even for the field-free case, in contrast with regular solutions of first-order Dirac equation. From the physical point of view, it can be seen as extending the phenomenological model a bit closer to the field approach. The latter one involves solutions of an infinite number of regular Dirac equations; they are used to describe virtual fermions and antifermions that screen bare charges. Since a nonzero pseudoscalar corresponds to general states which have both fermion and antifermion components, we attempt to effectively take into account the existence of such a screening cloud around real fermions while staying within the single particle approach. The additional parameter, the Dirac pseudoscalar, exists in the regular formalism; however, it has not been used in the conventional phenomenological models based on the prevailing assumption that free fermions do not possess nonzero pseudoscalar. We have developed the model that does not rely on this assumption and allows to test it experimentally.

The application of WKB approximation to the squared Dirac equation upgraded with the AMM and EDM terms has allowed to extend the phenomenological model of spin dynamics with the static pseudoscalar correction  \cite{porshnev_electron_2020, baryshevsky_search_2020}. Before reviewing and extending the WKB approach however, let us next briefly review the conventional model and discuss how the dynamically induced pseudoscalar is included into the regular formalism.

\section{Regular and squared Dirac equations with AMM and EDM terms}
In the minimal coupling, the standard Dirac formalism is based on the following equation 
\begin{equation}\label{eq20}
	\qty(i \slashed \partial -  e \slashed A-m) \psi = 0\,, 
\end{equation}
where $A_\mu$ is the electromagnetic potential. Several gauge-invariant bilinears can be constructed from a single bispinor \cite[p. 102]{berestetskii_quantum_1982} 
\begin{gather}
\begin{aligned}\label{relQM:eq59}
	&r &&= \bar{\psi}\psi&&=\abs{\phi}^2 - \abs{\chi}^2\,,	\\[1ex]
	&q &&= i\bar{\psi}\gamma^5\psi&&=i(\phi^*\chi - \chi^*\phi)\,,\\[1ex]
	&k^\mu &&= \bar{\psi}\gamma^\mu\psi&&=(\abs{\phi}^2 + \abs{\chi}^2,\,\,  \phi^*\bm{\sigma}\chi + \chi^*\bm{\sigma}\phi)\,,\\[1ex]
	&w^\mu &&= \bar{\psi}\gamma^\mu\gamma^5\psi&&=(\phi^*\chi + \chi^*\phi,\,\, \phi^*\bm{\sigma}\phi + \chi^*\bm{\sigma}\chi)\,,
\end{aligned}
\end{gather}
where their spinor representation is also given. They are real-valued quantities  with distinct transformation properties.
The products of bilinears are interrelated by the Fierz identities
\begin{gather}
\begin{aligned}\label{chFields:eq976}
	&k^\mu k_\mu = -w^\mu w_\mu &&= r^2 + q^2=n^2\, , &k^\mu w_\mu 				&&= 0\, , \\[1ex]
	&k^\mu w^\nu - w^\mu k^\nu &&= q B^{\mu\nu}+ r \tilde B^{\mu\nu}\, .
\end{aligned} 
\end{gather} 
where  the dual tensor is defined as
\begin{equation}
		\tilde B^{\mu\nu}= \frac{1}{2}\varepsilon^{\mu\nu\rho\sigma} B_{\rho\sigma} = -i\bar{\psi} \gamma^5 \sigma^{\mu\nu}\psi\,.
\end{equation}
Many more similar identities can be found. For example, we can reverse the last identity in \eqref{chFields:eq976} to obtain
\begin{equation}\label{eq75}
	\tilde B_{\mu\nu} = \frac{r(k_\mu w_\nu - w_\mu k_\nu) + q \varepsilon_{\mu\nu\rho\sigma} k^\rho w^\sigma}{r^2+q^2}\,.
\end{equation}
They are especially simple if the pseudoscalar $q$ is zero.

Two conservation laws can be derived from \eqref{eq20}
\begin{align}
	&\partial_\mu k^\mu =0\,,		&\partial_\mu w^\mu = 2m q\,,
\end{align}
where only the gauge-invariant Dirac bilinears are used.  However the equations of motion for other bilinears  cannot be given in the closed form by using only the gauge-invariant Dirac bilinears, see \cite{inglis_fierz_2014}.

The Dirac equation \eqref{eq20} cannot provide a finer description of atomic effects or spin dynamics, since it is the single particle equation. To improve its accuracy, it is commonly upgraded \cite{foldy_electromagnetic_1952,sokolov_relativistic_1974, bagrov_dirac_2014}  by adding two field-dependent terms 
\begin{equation}\label{eq2}
	\qty[i \slashed \partial -  e \slashed A-m - (\frac{a e}{2m}+id \gamma^5)\frac{\sigma^{\mu\nu}}{2}F_{\mu\nu}] \psi^\prime = 0\,, 
\end{equation}
where $a$ and $d$ are the anomalous magnetic and electric dipole moments. Hence, the minimal coupling is extended by adding two terms stemming from the additional fermion form-factors. Effectively, it is an attempt to implicitly take into account the radiative corrections \cite[p. 152]{berestetskii_quantum_1982} which consistent description is given only within the standard theory. The connection between the term with anomalous magnetic moment and the radiation field was shown in \cite{baier_radiative_1972}.  The definition \eqref{relQM:eq59} of bilinears does not change for the extended coupling scheme. 
The current conservation also stays the same, however the conservation equation for axial current does change
\begin{align}\label{eq100}
	&\partial_\mu k^{\prime\mu} =0\,,		&\partial_\mu w^{\prime\mu} = 2m q^\prime + \bar{\psi}^\prime\qty(\frac{a e}{2m} i\gamma^5- d)\sigma^{\mu\nu} \psi^\prime F_{\mu\nu}\,.
\end{align}
One can see a certain analogy here with the chiral anomaly that appears in QFT upon regularization of triangle diagrams \cite[p. 376]{ryder_quantum_1986}. 

It is convenient to eliminate $\gamma^5$ from \eqref{eq2} by introducing the dual field tensor
\begin{equation}\label{eq40}
	\qty(i \slashed \partial -  e \slashed A-m - \frac{1}{2}b_{\mu\nu}\sigma^{\mu\nu}) \psi = 0\,, 
\end{equation}
where we dropped the primes and defined the new tensor coefficient as
\begin{equation}\label{eq69}
	b_{\mu\nu} = \frac{a e}{2m}F_{\mu\nu}-d\tilde F_{\mu\nu}\,.
\end{equation}
Our next step is to obtain the second-order differential equation. Following the traditional approach \cite[p.  119]{berestetskii_quantum_1982}, it is achieved by squaring \eqref{eq40} with the adjoint operator
\begin{equation}\label{eq105}
	i \slashed \partial -  e \slashed A_\mu+m + \frac{1}{2}b_{\mu\nu}\sigma^{\mu\nu}\,. 
\end{equation}
If \eqref{eq105} is combined with the regular Dirac operator $(i\slashed \partial-m)$, while ignoring both potential $A_\mu$ and Pauli terms, the Klein-Gordon equation is obtained. The Pauli term in the  adjoint operator must change its sign \cite{rafanelli_classical_1964}, otherwise its linear contribution drops out. 
Dropping the terms with gradients and squares of field tensors, 
the squared Dirac equation is obtained as
\begin{equation}\label{chFields:eq1971}
	\Big[(i\partial_\mu-eA_\mu)^2 -m^2- ma_{\mu\nu} \sigma^{\mu\nu} -2ib_{\mu\nu} \gamma^\nu(i\partial^\mu-eA^\mu) \Big]\psi=0\,.
\end{equation}
where we defined one more tensor coefficient as
\begin{equation}\label{eq70}
	a_{\mu\nu} = \frac{e}{2m} F_{\mu\nu}+b_{\mu\nu} = \frac{(1+a) e}{2m}F_{\mu\nu}-d\tilde F_{\mu\nu}=\frac{g e}{4m}F_{\mu\nu}-d\tilde F_{\mu\nu}
\end{equation}
The equation \eqref{chFields:eq1971} can be seen as equivalent to two original Dirac equations in the limit of small and slow-changing electromagnetic fields. The spin-dependent terms are explicit now, since the kinetic term is scalar. 

\section{Dynamically induced pseudoscalar and Pauli equation with AMM and EDM terms}\label{dynamo}
We briefly review here the derivation of spin motion equation following the conventional approach that is applied to the regular Dirac equation with the AMM and EDM terms. The nonrelativistic limit for the standard representation  is obtained by eliminating the lower spinor, see \cite[p. 32]{akhiezer_quantum_1981} or \cite[p. 122]{berestetskii_quantum_1982}. Applied correctly, the reduced Hamiltonian is the same as obtained by the FW transformation \cite{de_vries_non-relativistic_1968}. 

In the non-relativistic limit ($v\to 0$) and in the standard representation, the lower spinor is expected to be much smaller than the upper one
	\begin{equation}\label{eq1}
		\psi^\prime = \mqty(\phi^\prime\\ \chi^\prime)\qquad\overset{v\to 0}{\to}\qquad \psi= \mqty(\phi\\ \chi)\,,
	\end{equation}
here $v$ is the three-velocity of fermion, and $\abs{\chi}\ll \abs{\phi}$. Excluding the rest energy with the   substitution
\begin{equation}
	\psi = \mqty(\phi\\ \chi)\, e^{-i m t}\,, 
\end{equation}
the equation \eqref{eq2} turns  into two linked spinor equations
\begin{gather}
\begin{aligned}\label{eq6}
	&\qty(i\pdv{}{t}-eA_0 + \bm{\sigma} \, \bm{\alpha}  )\phi && =- \bm{\sigma}\qty(i\bm{\partial}+ e\textbf{A} -i \bm{\kappa} )\chi\,,\\[1ex]
	&\qty(i\pdv{}{t}-eA_0 - \bm{\sigma} \, \bm{\alpha} +2m)\chi && =- \bm{\sigma}\qty(i\bm{\partial}+ e\textbf{A} +i   \bm{\kappa})\phi\,.
\end{aligned}
\end{gather}
where $\bm{\sigma} \bm{\alpha}=\sigma_i \alpha_i$, and two field-dependent three-vectors are defined as
\begin{align}\label{eq77}
	&\bm{\alpha} = d \,\textbf{E}+\frac{a e}{2m}\, \textbf{B}\,,	&\bm{\kappa} = \frac{a e}{2m} \textbf{E}-d\, \textbf{B}\,.
\end{align}
The cross terms $d \textbf{B}$ and $a\textbf{E}$ are originated by the term $\gamma^5\sigma^{\mu\nu}$ in the  Dirac equation \eqref{eq2}. They will lead to corresponding vector products of fields with momentum in the spin equations below.  In the first approximation, we retain only the term $2m\chi$ in the left-hand side of second equation in \eqref{eq6} to obtain
\begin{equation}\label{eq4}
	\chi = -\frac{1}{2m}\bm{\sigma}\qty(i\bm{\partial}+ e\textbf{A} +i \bm{\kappa} )\phi\,.
\end{equation}
To simplify derivations, we assume the weak field approximation for which both electric and magnetic fields $(E_i, H_i)$ are considered nearly constant and small, while the potential $A_\mu(x)$ does vary with space and time. Accordingly, any products bilinear in fields  will be ignored.

Since $\bm{\kappa}=0$ in the absence of fields, the expression \eqref{eq4} shows that the bottom spinor is zero for constant upper spinor $\phi$. The pseudoscalar density is obtained as
\begin{multline}\label{eq5}
	2m\, q_{dyn} = 2m i\bar{\psi}\gamma^5\psi = 2m i (\phi^*\chi - \chi^*\phi) = \phi^*\sigma_i (\partial_i\phi)-i\phi^*\sigma_i\qty( eA_i +i \kappa_i)\phi\\[1ex]
	+(\partial_i\phi^*)\sigma_i \phi+ i\phi^*\sigma_i\qty(e A_i -i \kappa_i)\phi = \partial_i(\phi^*\sigma_i\phi)+ 2 \kappa_i\,(\phi^*\sigma_i\phi)\,.
\end{multline} 
Ignoring the term with dipole moments, \eqref{eq5} acquires the well known form \cite[p. 16]{khriplovich_cp_1997} which determines the divergence of spin density $\phi^*\sigma_i\phi$ in the non-relativistic case. We see again that in the conventional model the pseudoscalar is zero  for constant upper spinor, or equivalently, in the absence of external fields, per the  above discussion of conventional models. The current density $\textbf{j}=\phi^*\bm{\sigma}\chi + \chi^*\bm{\sigma}\phi$ also disappears if the lower spinor is zero.  Under external fields, the lower spinor, hence both pseudoscalar and current too, evolves, and its dynamic evolution is included into the equation for upper spinor. The addition of AMM and EDM terms to the regular Dirac equation does not change this conclusion, since $\bm{\kappa}$ is still field-dependent.

As an example, consider an electron orbital $\phi= \phi_s \Phi(x)$ for a case where it can be given as a product of spin eigenfunction $\phi_s$ and the space part $\Phi(x)$. It implies that the spin state is fixed for the entire orbital, and the spin projection number is well defined. The equation \eqref{eq5} is then manipulated as
\begin{equation}\label{eq102}
	2m\, q_{dyn} = \textbf{s}\cdot(\bm{\nabla} \rho + 2 \rho\bm{k})\,,
\end{equation}
where the spin vector is $\textbf{s} = \phi_s^*\bm{\sigma}\phi_s$, and the spinor density is $\rho = \Phi^*\Phi$. The dynamically induced pseudoscalar is non zero if the density gradient and $\bm{\kappa}$ are nonzero, even if the spin state $\phi_s$ is fixed.

Since we use the natural units, it might not be immediately clear that both terms in the right hand sides of \eqref{eq5} and \eqref{eq102} are proportional to the Planck constant $\hbar$. Hence, the dynamically induce pseudoscalar is of $\order{\hbar}$. This comment will be needed later on for a comparison with motion equations derived within the WKB approach. In the weak-field limit, the pseudoscalar is found to be a constant of motion to first order of the semiclassical WKB solution if we ignore the corrections of $\order{\hbar}$ which is consistent with \eqref{eq5} and \eqref{eq102}.

Next, substituting \eqref{eq4} into the first equation in \eqref{eq6}, we obtain
\begin{equation}\label{eq10}
	i\pdv{\phi}{t} = H^{(1)}\phi = \qty[\frac{1}{2m}\textbf{p}^2 +e A_0 -\frac{e}{2m}\bm{\sigma}\textbf{H} - \bm{\sigma}\bm{\alpha} +\frac{1}{m} \bm{\sigma} \big(\bm{\kappa}\cross \textbf{p}) ]\,,
\end{equation}
where the term bilinear in fields was dropped, and  $\textbf{p}=i\bm{\partial}+e\textbf{A}$. It is the Pauli-like equation that follows to first order from the non-minimally coupled Dirac equation \eqref{eq2} in the non-relativistic limit.

The next iteration requires a refinement of \eqref{eq4}, and leads to the well-known relativistic and field gradient corrections to Hamiltonian \eqref{eq10}. It will also require an additional non-unitary transformation to ensure the conservation of $\abs{\phi}^2$. Since the approach is well known \cite{akhiezer_quantum_1981,berestetskii_quantum_1982, de_vries_non-relativistic_1968, strange_relativistic_1998} and the role of dynamically induced pseudoscalar is clarified, the Hamiltonian to second order is given without derivation
\begin{multline}\label{eq78}
	\qquad\qquad H^{(2)} = \frac{\textbf{p}^2}{2m} +e A_0 -\frac{ge}{4m}\bm{\sigma}\textbf{B}- \qty(\frac{ae}{2m}+\frac{e}{4m})\bm{\sigma}\big( \textbf{E}\cross\frac{\textbf{p}}{m}\big)\\[1ex]
		- d\,\bm{\sigma}\textbf{E}-d\,\bm{\sigma}\big(\textbf{B}\cross \frac{\textbf{p}}{m}\big) \,,\qquad
\end{multline}
here we dropped the terms with higher powers of momentum, field gradients, and products of fields with second power of momentum, including the Darwin term. The apparent non-hermiticity of terms with vector product is not an issue, as discussed in \cite{khriplovich_equations_1999} and \cite[p. 51]{bjorken_relativistic_1964}. The form properly accounts for the Thomas precession which can also be obtained from the BMT equation \cite[sec.41]{berestetskii_quantum_1982}.

The operator equation of motion for spin  $\hat s = \bm{\sigma}/2$ is given per standard rules
\begin{multline}\label{eq80}
	\qquad\qquad\dv{\hat s}{t} = i[H^{(2)}, \hat s]= \frac{ge}{2m}(\hat s\cross \textbf{B})+\qty(\frac{ae}{m}+\frac{e}{2m})\big[\hat s\cross(\textbf{E}\cross \frac{\textbf{p}}{m})\big] \\[1ex]
	+ 2d \Big[ \hat s\cross\textbf{E}-\hat s\cross(\textbf{B}\cross \frac{\textbf{p}}{m})\Big] \,.
\end{multline}
The obtained Hamiltonian without the EDM part is solidly supported by innumerous atomic data, as we already mentioned before, see the discussion around equation \eqref{eq27} in the introduction. Having in mind its application to storage ring experiments \cite{mane_spin-polarized_2005}, we consider next  the  quasi-classical approximation. For this purpose, we average \eqref{eq80} over a wave packet with a sufficiently narrow spread, please see the discussion on applicability of such an approximation in \cite[sec.41]{berestetskii_quantum_1982} and \cite{baier_radiative_1972, mane_spin-polarized_2005}. The evolution of spin expectation value $\textbf{s} = \langle \hat s\rangle$ is then described as
\begin{multline}\label{eq79}
	\qquad\qquad\dv{\textbf{s}}{t} = \frac{ge}{2m}(\textbf{s}\cross \textbf{B})+\qty(\frac{ae}{m}+\frac{e}{2m})\big[\textbf{s}\cross(\textbf{E}\cross \textbf{v})\big]\\[1ex] + 2d \Big[ \textbf{s}\cross\textbf{E}-\textbf{s}\cross(\textbf{B}\cross \textbf{v})\Big] =  \textbf{s}\cross\Omega_{nr} \,,
\end{multline}
where $\textbf{v} = \langle\textbf{p}\rangle/m$, and $\Omega_{nr}$ is the nonrelativistic precession frequency
\begin{equation}\label{eq90}
	\Omega_{nr} = \frac{ge}{2m}\textbf{B}+\qty(\frac{ae}{m}+\frac{e}{2m})\textbf{E}\cross \textbf{v} + 2d \big( \textbf{E}-\textbf{B}\cross \textbf{v}\big)\,.
\end{equation}
In the semiclassical approximation, we assume that the space part of wave packet moves along an orbit that is  a solution of corresponding Hamilton-Jacobi equation. More accurately, such orbits can be described by eigenfunctions of scalar part of Hamiltonian, since particle trajectories are only weakly influenced by spin in this approximation. The spin evolution along such trajectories is given by \eqref{eq79}.  

For completeness, we also give the relativistic precession frequency \cite{nelson_search_1959,  berestetskii_quantum_1982, derbenev_polarization_1973, fukuyama_searching_2012}
\begin{multline}\label{eq83}
	\Omega = \frac{e}{m}\qty[\qty(a+\frac{1}{\gamma}) \textbf{B}-\frac{a\gamma}{\gamma+1}  (\textbf{v}\cdot\textbf{B}) \,\textbf{v}-\qty(a+\frac{1}{\gamma+1})\textbf{v}\cross\textbf{E} ]\\[1ex]
	 + 2d \qty[ \textbf{E} - \frac{\gamma}{\gamma+1} (\textbf{v}\cdot\textbf{E}) \,\textbf{v}+ \textbf{v}\cross \textbf{B} ]\,.
\end{multline}
where $\textbf{s}$ is given in the rest frame, while the fields are in the laboratory frame, and $\gamma$ is the Lorentz factor. This form is derived from the BMT equation \cite{berestetskii_quantum_1982, fukuyama_derivation_2013}, hence it should be valid for arbitrary velocities even if it is not written in a Lorentz covariant form. Expanding the Lorentz factor $\gamma$ in \eqref{eq83} from the one hand, and recovering the relativistic corrections in Hamiltonian \eqref{eq78} from the other hand, it is possible to show that two corresponding spin equations match each other at least up to the fourth power of momentum. The spin equations show that the magnetic anomaly couples differently to magnetic and electric fields in the laboratory frame \cite{mane_spin-polarized_2005} which is expected since the magnetic moment couples to only magnetic fields in the rest frame.

\section{WKB solution of squared Dirac equation}
Our goal here is to briefly review the previous results \cite{porshnev_electron_2020, baryshevsky_search_2020}, and extend them by obtaining the second order corrections. For this purpose, we recover the Planck constant in the squared Dirac equation \eqref{chFields:eq1971}
\begin{equation}\label{eq48}
	\Big[(i\hbar\partial_\mu-eA_\mu)^2 -m^2- m\hbar a_{\mu\nu} \sigma^{\mu\nu} -2i\hbar b_{\mu\nu} \gamma^\nu(i\hbar\partial^\mu-eA^\mu) \Big]\psi=0\,.
\end{equation}
Following \cite[p. 36]{akhiezer_quantum_1981}, the bispinor is approximated as
\begin{equation}\label{eq49}
	\psi = e^{i S/\hbar}\, f= e^{i S/\hbar}(f_0 + \hbar f_1 +\hbar^2 f_2+\dots)\,,
\end{equation}
where $S(x)$ and $f_i(x)$ are unknown scalar and bispinor-valued functions respectively. 
Upon substituting \eqref{eq49} into \eqref{eq48}, we obtain
\begin{gather}
\begin{aligned}\label{eq50}
	&p_\mu p^\mu - m^2&&= \phantom{-}0\,,\\[1ex]
	&\Big[2p_\mu \partial^\mu+ (\partial^\mu p_\mu)    +im a_{\mu\nu} \sigma^{\mu\nu}- 2 b_{\mu\nu}\gamma^\nu p^\mu\Big] f_0&&= \phantom{-}0\,,\\[1ex]
	&\Big[2p_\mu \partial^\mu+ (\partial^\mu p_\mu) +im a_{\mu\nu} \sigma^{\mu\nu}- 2 b_{\mu\nu}\gamma^\nu p^\mu\Big] f_1&&=-  i(\partial^2 f_0+2 b_{\mu\nu}\gamma^\nu \partial^\mu f_0)\,,\\[1ex]
	\dots
\end{aligned}
\end{gather}
where we define the four-vector $p_\mu$ as 
\begin{equation}\label{eq51}
	p_\mu = - (\partial_\mu S+eA_\mu)\,.
\end{equation}
The expression in \eqref{eq50} has the form of Hamilton-Jacobi equation where the phase derivatives $\partial_\mu S$ are associated with the classical Hamiltonian and conjugate momentum as in \eqref{eq51}.

The assignment \eqref{eq51} is straightforward and aligned with the standard way of taking the classical limit in the Dirac formalism. However to make it self-consistent within the WKB approach, several other things must fall into their places. Since the Hamilton-Jacobi equation  is  one of many equivalent forms that describe the classical motion, the traditional four-dimensional equation of motion follows. 
However, we will see that the motion equation can also be derived from the second equation in \eqref{eq50} if the assignment \eqref{eq51} is extended to include the four-velocity $u_\mu$
\begin{equation}\label{eq53}
	p_\mu = - (\partial_\mu S+eA_\mu) = m u_\mu\,,
\end{equation}
where $u_\mu u^\mu =1$.  First, interpreting the first equation in \eqref{eq50} as classical Hamilton-Jacobi equation together with \eqref{eq51} means that particle trajectories are not influenced by their spin in this quasi-classical limit. Second, the extended assignment \eqref{eq53} forces the momentum to be parallel to velocity. Hence, this choice eschew models where the momentum is not parallel to velocity which are accompanied by rapid oscillating motion or zitterbewegung \cite{stachel_classical_1977}. Ensuring also that only the Dirac bilinears from table \ref{eedm_table} are used, the motion equations will be covariant, gauge-invariant and free of potentially unwanted artifacts.  

The expectation values of momentum and spin are linked with Dirac bilinears as
\begin{align}\label{eq55}
	&p_\mu=\frac{m}{n}\, \bar{f}\gamma_\mu f\,, 
	&s_\mu= \frac{1}{n}\,  \bar{f}\gamma_\mu\gamma^5 f\,,
\end{align}
where they normalized with quantity $n=\sqrtP{r^2+q^2}$ per Fierz identities, and $s_\mu s^\mu = -1$. The corresponding approximation to first order in $\hbar$  follows as
\begin{equation}
	r = \bar{f}{f} = (\bar{f}_0 + \hbar \bar{f}_1+\dots)(f_0+\hbar f_1+\dots) = \bar{f}_0 f_0 + \hbar\qty(\bar{f}_0 f_1 + \bar{f}_1 f_0)+\order{\hbar^2}\,,
\end{equation}
and analogously for other bilinears.
The original current conservation is stated as 
\begin{equation}\label{chFields:eq1975}
	\partial_\mu(\bar{\psi}\gamma^\mu\psi)= \frac{1}{m}\partial_\mu(n p^\mu)= \partial_\mu(n u^\mu)=0\,,
\end{equation}
which simplifies to 
\begin{equation}\label{eq61}
	\partial_\mu(r u^\mu)=0+\order{q^2/r^2}\,,
\end{equation}
if the pseudoscalar $q=i\bar{f}\gamma^5f$ turns zero. The total (substantive) derivative over proper time is defined as \cite{rafanelli_classical_1964}
\begin{equation}\label{eq56}
	\dv{}{\tau} =u_\mu \partial^\mu = \frac{1}{m }p_\mu \partial^\mu  \,.
\end{equation}
Combining \eqref{chFields:eq1975} with \eqref{eq56}, we see that the change in some quantity $\Phi$ over particle trajectory
\begin{equation}\label{eq57}
	n\dv{\Phi}{\tau}= n u_\mu \partial^\mu \Phi = \partial^\mu( n  u_\mu \Phi) - \Phi \underbrace{\partial_\mu(n u^\mu)}_0 = \partial^\mu( n  u_\mu \Phi)
\end{equation}
is equivalent to the divergence of its flux along particle trajectories in this model.
\\[1ex]

\noindent\textbf{Density: } The equation for $r=\bar{f}f$ follows from the system \eqref{eq50} as
\begin{equation}\label{eq58}
	\partial_\mu (r p^\mu)
	 = 2 b_{\mu\nu} p^\mu \,\bar{f}\gamma^\nu f+ \hbar C_r
	  = 2 \frac{n}{m} \underbrace{b_{\mu\nu} p^\mu p^\nu}_0+ \hbar C_r = \order{\hbar} \,,
\end{equation}
where the correction term which is first order in $\hbar$ is given as
\begin{equation}
		C_r = \frac{i}{2} \qty[(\partial^2 \bar{f}_0) f_0 - \bar{f}_0 (\partial^2  f_0)]+ i b_{\mu\nu}\qty[(\partial^\mu \bar{f}_0) \gamma^\nu f_0 -\bar{f}_0\gamma^\nu (\partial^\mu f_0 )]+ \order{\hbar}\,.
\end{equation}
This is where the assignments \eqref{eq53} and \eqref{eq55} make the leading (classical) term vanish \cite{rafanelli_classical_1964}. The non-vanishing term which forces the ratio $r/n$ to change along its trajectory is proportional to Planck constant. Using \eqref{eq57}, the equation \eqref{eq58} is re-written as
\begin{equation}\label{eq59}
	\partial_\mu (r p^\mu) = m \partial_\mu (r u^\mu) = m n \dv{}{\tau}\qty(\frac{r}{n}) =  \hbar C_r\,,
\end{equation}
For $\abs{q}\ll r$,
the total derivative in \eqref{eq59} turns   into 
\begin{equation}\label{eq60}
	 m n \dv{}{\tau}\qty(\frac{r}{n}) = -m\frac{r}{2} \dv{}{\tau}\qty(\frac{q}{r})^2+\dots=  \hbar C_r\,,
\end{equation}
which means that a change in pseudoscalar along particle trajectories in this model is the quantum correction, since it is proportional to $\hbar$.  

Concluding this part, the quasiclassical particle density is a constant of motion to first order
\begin{equation}\label{eq65}
	\dv{r}{\tau} = 0 + \order{\hbar}\,,
\end{equation}
the result that was already anticipated from \eqref{eq61}. 
\\[1ex]

\noindent\textbf{Pseudodensity: }
The equation for $q=i\bar{f}\gamma^5f$ follows from the system \eqref{eq50} as
\begin{equation}\label{eq62}
	\partial_\mu ( q  p^\mu)
	 = 0+  \hbar C_q  \,,
\end{equation}
where the correction term which is first order in $\hbar$ is given as
\begin{equation}
		C_q  =  \frac{1}{2} \qty[\bar{f}_0 \gamma^5(\partial^2  f_0)-(\partial^2 \bar{f}_0)\gamma^5 f_0 ]- b_{\mu\nu}\partial^\mu (n s^\nu)+ \order{\hbar}\,.
\end{equation}
Using \eqref{eq57}, the equation \eqref{eq62} is re-written as
\begin{equation}\label{eq63}
	\partial_\mu (q p^\mu) = m \partial_\mu (q u^\mu) = m n \dv{}{\tau}\qty(\frac{q}{n}) =  \hbar C_q\,,
\end{equation}
For $\abs{q}\ll r$,
 the ratio is expanded as
\begin{equation}
	\frac{q}{n} =\frac{q}{r} - \frac{1}{2}\qty(\frac{q}{r})^3+\dots\,,
\end{equation}
the total derivative in \eqref{eq63}  turns  into 
\begin{equation}\label{eq64}
	 m n \dv{}{\tau}\qty(\frac{q}{n}) =  mr\dv{}{\tau}\qty(\frac{q}{r})+\dots=  \hbar C_q\,,
\end{equation}
which is consistent with the previous conclusion: a change in pseudoscalar is the quantum correction in this model.  

Using the approximation \eqref{eq65}, the equation \eqref{eq64} becomes 
\begin{equation}\label{eq103}
	\dv{q}{\tau} = 0 + \order{\hbar}\,,
\end{equation}
which means that both densities are constants of motion to first order in the weak field approximation. As we have already mentioned, the dynamically induced part of pseudoscalar is of $\order{\hbar}$, see the discussion in section \ref{dynamo}, which is consistent with the equation \eqref{eq103}.
\\[1ex]

\noindent\textbf{Four-velocity: }
The equation for $n u_\mu=\bar{f}\gamma_\mu f$ follows from the system \eqref{eq50} as
\begin{equation}\label{eq66}
	\partial_\mu ( n u^\rho  p^\mu) = 2 m a^{\rho\nu} n u_\nu +2r  b^{\mu\rho} p_\mu+ \hbar C_u  \,,
\end{equation}
where the correction term which is first order in $\hbar$ is given as
\begin{multline}
		\qquad\qquad C_u  = \frac{i}{2} \qty[(\partial^2 \bar{f}_0)\gamma^\rho f_0 - \bar{f}_0 \gamma^\rho(\partial^2  f_0)]\\[1ex]
		+ i b_{\mu\nu}\qty[(\partial^\mu \bar{f}_0) \gamma^\nu\gamma^\rho f_0 -\bar{f}_0\gamma^\rho\gamma^\nu (\partial^\mu f_0 )]+ \order{\hbar}\qquad\qquad \,.
\end{multline}
Using \eqref{eq57}, the equation \eqref{eq58} is re-written as
\begin{equation}\label{eq67}
	\partial_\mu ( n u^\rho  p^\mu) = m n \dv{u^\rho}{\tau} = 2 m a^{\rho\nu} n u_\nu +2r  b^{\mu\rho} p_\mu+ \hbar C_u\,,
\end{equation}
which leads to the following equation 
\begin{multline}
	\dv{u^\rho}{\tau} = 2 a^{\rho\nu} u_\nu + 2\frac{r}{n} b^{\mu\rho} u_\mu + \order{\hbar}
	=2 \qty(\frac{e}{2m} F^{\rho\nu}+b^{\rho\nu}) u_\nu - 2\frac{r}{n} b^{\rho\nu} u_\nu + \order{\hbar} \\[1ex]
	= \frac{e}{m} F^{\rho\nu} u_\nu + 2\qty(1-\frac{r}{n}) b^{\rho\nu} u_\nu+ \order{\hbar}= \frac{e}{m} F^{\rho\nu} u_\nu + \frac{q^2}{r^2}b^{\rho\nu} u_\nu + \order{\hbar}\,,
\end{multline}
where we assumed $\abs{q}\ll r$ and used the definitions of tensor coefficients from \eqref{eq69} and \eqref{eq70}. Ignoring two last terms which should be small in the quasiclassical limit, we obtain the classical equation of motion 
\begin{equation}
	\dv{u^\rho}{\tau}=\frac{e}{m} F^{\rho\nu} u_\nu\,,
\end{equation}
which shows the consistency within the WKB approximation \cite{rafanelli_classical_1964}. 
\\[1ex]

\noindent\textbf{Spin: }
The equation for $n s_\mu=\bar{f}\gamma_\mu \gamma^5f$ follows from the system \eqref{eq50} as
\begin{equation}\label{eq72}
	\partial_\mu ( n s^\rho  p^\mu) =  2 m a^{\rho\nu} n s_\nu +2 i b_{\mu\nu} p^\mu\bar{f}\sigma^{\rho\nu}\gamma^5 f+ \hbar C_s  \,,
\end{equation}
where the correction term which is of first order in $\hbar$ is given as
\begin{multline}
		\qquad\qquad C_s  =\frac{i}{2} \qty[(\partial^2 \bar{f}_0)\gamma^\rho\gamma^5 f_0 - \bar{f}_0 \gamma^\rho\gamma^5(\partial^2  f_0)]\\[1ex]
		+ i b_{\mu\nu}\qty[(\partial^\mu \bar{f}_0) \gamma^\nu\gamma^\rho\gamma^5 f_0 -\bar{f}_0\gamma^\rho\gamma^5\gamma^\nu (\partial^\mu f_0 )]+ \order{\hbar}\textbf{}\,.
\end{multline}
The terms  of second order in $\hbar$ are not shown. Using \eqref{eq57}, the spin motion equation \eqref{eq58} is re-written as
\begin{equation}\label{eq73}
	\dv{s^\rho}{\tau} =  2  a^{\rho\nu}  s_\nu +\frac{2 }{n}i b_{\mu\nu} u^\mu\bar{f}\sigma^{\rho\nu}\gamma^5 f+ \frac{\hbar}{mn} C_s\,.
\end{equation}
The second term is further manipulated as
\begin{multline}
 b_{\mu\nu} u^\mu(i\bar{f}\sigma^{\rho\nu}\gamma^5 f) =- b_{\mu\nu} u^\mu \tilde B^{\rho\nu}=- b_{\mu\nu} u^\mu \big[r(u^\rho s^\nu - s^\rho u^\nu) + q \varepsilon^{\rho\nu\lambda\sigma} u_\lambda s_\sigma\big]\\[1ex]
 = r s^\mu b_{\mu\nu} u^\nu u^\rho+ q \tilde b^{\rho\nu} s _\nu + q s^\mu \tilde b_{\mu\nu} u^\nu u^\rho\,,
\end{multline}
where the Fierz identity \eqref{eq75} is used. It leads to the following equation 
\begin{multline}
	\dv{s^\rho}{\tau} = \qty(\frac{g e}{2m}+2d \frac{q}{n}) F^{\rho\nu}  s_\nu +  \qty(\frac{r}{n}  \frac{a e}{m}+  2d \frac{q}{n})s^\mu F_{\mu\nu} u^\nu u^\rho\\[1ex]
	 -\qty( 2d - \frac{a e}{m}\frac{q}{n} ) \tilde F^{\rho\nu}  s_\nu -  \qty( 2d \frac{r}{n} -  \frac{a e}{m}\frac{q}{n})s^\mu \tilde F_{\mu\nu} u^\nu u^\rho+ \frac{\hbar}{mn} C_s \,,
\end{multline}
where we used the tensor coefficients from \eqref{eq69} and \eqref{eq70}. Since $\abs{q}\ll r$, the density ratios can be expanded to yield
\begin{multline}\label{eq76}
	\qquad\qquad \dv{s^\rho}{\tau} = \frac{g_1 e}{2m} F^{\rho\nu}  s_\nu + \frac{ e a_2 }{m}s^\mu F_{\mu\nu} u^\nu u^\rho\\[1ex]
	 -2 d^\prime \qty(\tilde F^{\rho\nu}  s_\nu +  s^\mu \tilde F_{\mu\nu} u^\nu u^\rho)+ \frac{\hbar}{m r} C_s\,,\qquad\qquad 
\end{multline}
where the terms with higher powers of $q$ were dropped, and the effective moments are defined as
\begin{gather}
\begin{aligned}\label{eq89}
	&g_1 &&= 2(1+a_1)\,,			&&\qquad\qquad a_1 &&= a + \frac{2 d m}{e}\, \beta\,,\\[1ex]
	&d^\prime &&= d -  \frac{a e}{2m}\, \beta\,,	&&\qquad\qquad  a_2 &&= a\qty(1-\frac{\beta^2}{2}) + \frac{2 d m}{e}\, \beta\,,
\end{aligned}
\end{gather}
here $\beta = q /r$. Since $\abs{d}\ll \abs{ae/m}$ and $\abs{\beta}\ll 1$, the corrections to $a$ are of second order of smallness.  

Inspecting the effective coefficients in \eqref{eq89}, we see that the pseudoscalar mixes the intrinsic magnetic and electric moments in them. It is similar to $T$-odd  and $P$-odd atomic polarizabilities that are originated by symmetry-violating interactions of atomic electrons with nucleus nucleons   \cite{baryshevsky_high-energy_2012}. The scalar parts of these polarizabilities are pseudoscalar quantities which mix electric and magnetic contributions into the system energy. It is a quite typical situation in systems where symmetry-violating interactions are allowed, as we discussed in \cite{baryshevsky_search_2020}. 

The parameter $\beta$ is the ratio of fermion pseudodensity $q=i\bar\psi\gamma^5\psi$ to its regular $CP$-invariant density $r=\bar\psi\psi$. The first order WKB approximation simply says that two densities are constants of motion that are not influenced by four-velocity, spin or weak EM fields
\begin{align}
	&\dv{r}{\tau} = 0 + \order{\hbar}\,,			&\dv{q}{\tau} = 0 + \order{\hbar}\,.
\end{align}
The next approximation order in $\hbar$ will yield their dynamic parts which are induced by external fields, per agreement with the discussion in section \ref{dynamo}. Since we expect that for semiclassical or quasi-free motion the static part of regular density $r_0\approx 1$ should be close to one, it leaves completely open what is the static part of pseudodensity $q_0$. Saying otherwise, what is the value of $\beta_0$ for free fermions? Its physical meaning is the fraction of $CP$-noninvariant condensate relative to the $CP$-invariant one; similarly, it is the fraction of antifermions around a fermion that screens its bare charge. The proposed phenomenological model allows to develop ways for its experimental determination.

While the original AMM and EDM coefficients receive the pseudoscalar corrections, the functional form of spin equation (the dependencies on field tensor and four-velocity) in the first order of WKB approximation remains exactly the same as for the original BMT equation . The new spin equation collapses to the latter one if $\beta=0$. Since the pseudoscalar is the well-behaved gauge-invariant Dirac bilinear, the equation \eqref{eq76} is self-consistent, Lorentz-covariant, gauge-invariant, and free of artifacts, like zitterbewegung. It satisfies the first set of requirements for an extension of phenomenological description. Second, the corrections to magnetic anomaly are very small, since they are proportional to either $\beta^2$ or $d \beta$, where both factors are expected to be small individually. Instead, the correction to EDM is proportional to $ a\beta $, and might be significant since the intrinsic EDM is very small itself. Hence, the extended model seems to satisfy the stringiest requirements that we discussed in the introduction. 

The remaining task for this section is to re-write the spin equation for the laboratory frame. Since it has the same functional form as the original BMT equation, the known derivations \cite{berestetskii_quantum_1982, fukuyama_derivation_2013} can be directly used.  We only need to keep track of new coefficients which themselves are Lorentz-invariant. The precession frequency $\Omega_q$ with the pseudoscalar corrections is given as
\begin{multline}\label{eq86}
	\qquad\qquad\Omega_q =\frac{e}{m} \Big[\big(a^\prime+\frac{1}{\gamma}\big) \textbf{B}-\frac{a^{\prime\prime\prime}\gamma}{\gamma+1}  (\textbf{v}\cdot\textbf{B}) \,\textbf{v}-\big(a^{\prime\prime}+\frac{1}{\gamma+1}\big)\textbf{v}\cross\textbf{E}) \Big]\\[1ex]
		 + 2d^\prime \qty[ \textbf{E} - \frac{\gamma}{\gamma+1} (\textbf{v}\cdot\textbf{E}) \,\textbf{v}+ \textbf{v}\cross \textbf{B} ]\,,\qquad
\end{multline}
where the primed magnetic coefficients are defined as
\begin{gather}
\begin{aligned}\label{eq94}
	&a^\prime &&= a + 2d \frac{m}{e}\beta - a \frac{\beta^2}{2}\frac{\gamma-1}{\gamma}\,,\\[1ex]
	&a^{\prime\prime}&&=a + 2d \frac{m}{e}\beta - a \frac{\beta^2}{2}\frac{\gamma}{\gamma+1}\,,\\[1ex]
	&a^{\prime\prime\prime}&& = a + 2d \frac{m}{e}\beta-  a \frac{\beta^2}{2}\,,
\end{aligned}
\end{gather}
while $d^\prime$ is still given by \eqref{eq89}. All magnetic coefficients tend to the same value at ultrarelativistic velocities $\gamma\gg 1$
\begin{equation}	
	a^\prime = a^{\prime\prime} = a^{\prime\prime\prime}\,,
\end{equation}
and they all turn to $a$ at any velocity if $q=0$. The nonrelativistic precession frequency follows as
\begin{equation}\label{eq87}
	\Omega_q = \frac{e}{m} \Big[\big(a^\prime+1\big) \textbf{B}-\big(a^{\prime\prime}+\frac{1}{2}\big)\textbf{v}\cross\textbf{E}) \Big]\\[1ex]
			 + 2d^\prime \qty[ \textbf{E} + \textbf{v}\cross \textbf{B} ]+ \order{v^2}\,.
\end{equation}
where we dropped terms with powers of velocity equal or higher than two. 

Concluding, we derived the equation of spin motion within the WKB approximation to the squared Dirac equation.  Ignoring the pseudoscalar corrections, it matches the conventional BMT equation with added AMM and EDM terms. The pseudoscalar corrections to magnetic moments are of second order of smallness, while the correction to the EDM coefficient might be significant.  We also found that the functional form of spin equation in the covariant form \eqref{eq86} is the same as for the BMT equation; only the AMM and EDM coefficients acquire the Lorentz-invariant corrections. Hence, we suggest extending our results to other phenomenological models which use similar coefficients in their description of experimental data. For such a purpose, the conventional coefficients $a$ and $d$ must be replaced with expressions from \eqref{eq89}.

\section{Discussion}
Let us compare two versions of spin equation for the non-relativistic case. The conventional precession frequency \eqref{eq90} which is obtained from the Dirac equation by means of either the method of elimination or the FW transformation is given as
\begin{equation}\label{eq82}
	\Omega_{nr} = \frac{e}{m}\qty[(a^{exp}+1)\textbf{B}+\qty(a^{exp}+\frac{1}{2})\textbf{E}\cross \textbf{v}] + 2d^{exp} \big( \textbf{E}-\textbf{B}\cross \textbf{v}\big)\,.
\end{equation}
where the superscript is added to both $a$  and $d$. It highlights that these coefficients are inferred from applying this phenomenological model to experimental set-ups. Assuming that the model adequately describe experiments, they are matched versus theoretically evaluated ones
\begin{align}\label{eq92}
	&a^{exp} = a^{th}\,,			&d^{exp} = d^{th}\,.
\end{align}
A mismatch is interpreted as a lack of sufficient experimental precision, theoretical accuracy, or a potential indication of new physics.

The alternative spin equation is derived from the squared Dirac equation, and the corresponding precession frequency \eqref{eq87} has the same functional form 
\begin{equation}\label{eq91}
	\Omega_q = \frac{e}{m} \Big[\big(a^{exp}+1\big) \textbf{B}-\big(a^{*exp}+\frac{1}{2}\big)\textbf{v}\cross\textbf{E}) \Big]\\[1ex]
			 + 2d^{exp} \qty[ \textbf{E}-\textbf{B}\cross \textbf{v} ]\,.
\end{equation}
where again the experimental data are used to extract both $a^{exp}$ and $d^{exp}$ which are coefficients in front of corresponding terms in the spin equation \eqref{eq91}. However now, the connection with the theoretical values is given by expressions \eqref{eq94} which for the nonrelativistic case are
\begin{gather}
\begin{aligned}\label{eq93}
	&a^{exp} &&= a^{th} + 2d^{th} \frac{m}{e}\beta \,,\\[1ex]
	&a^{*exp}&&=a^{th}+ 2d^{th} \frac{m}{e}\beta-a^{th}\frac{\beta^2}{4} \,,\\[1ex]
	&d^{exp}&& = d^{th} -  \frac{a^{th} e}{2m}\beta\,.
\end{aligned}
\end{gather}
If the experimental data are treated with the conventional model \eqref{eq82} and $\beta\ne 0$, then we expect to see a difference between experimentally measured $a^{exp}$ and $d^{exp}$ and their values predicted by field theories. 

Since the pseudoscalar is the constant of motion in this approximation (quasi-free motion in the weak-field limit), as we discussed before, the choice between two versions is decided by an answer to the following question: what is the value of fermion pseudoscalar $q_0$ at rest? Saying otherwise, does the fermion condensate has a tiny $CP$-noninvariant fraction? It is implicitly or explicitly set to zero in conventional models.  Our main hypothesis is that this parameter can be nonzero for free fermions. The hypothesis cannot be rejected on purely logical grounds since the $P$- and $T$-odd quantities are admitted into the standard theory. Besides insisting on parity and time-reversal symmetries, there is no fundamental law that prohibits nonzero pseudoscalar for free fermions. 

We already mentioned that the virtual fermion-antifermion pairs around a fermion can be responsible for a nonzero static pseudoscalar. Some additional mechanisms can also be mentioned, though a discussion regarding specific mechanisms of generating static pseudoscalar is outside the scope of this work. The interaction of fermions with background pseudoscalar fields (axions) is the potential candidate. In addition to resolving the strong CP problem, it has been discussed as one of the possible mechanisms of generating a time-dependent EDM \cite{graham_axion_2011,stadnik_axion-induced_2014, budker_proposal_2014, abel_search_2017}.  Since such an interaction can generate the intrinsic EDMs, it could also be responsible for giving fermions a small pseudoscalar, though it is expected to be time-varying. Additionally, we have already mentioned the chiral anomaly \cite[p. 376]{ryder_quantum_1986} in discussing the equation for axial current divergence \eqref{eq100}. Introducing the AMM and EDM terms into the Dirac equation adds an extra term to this equation, see \eqref{eq100}. Since the pseudoscalar is the source of axial current, breaking its conservation could also lead to appearance of nonzero pseudoscalar. We postpone the in-depth discussion of this topic to some other time. Whatever a specific mechanism of generating a CP-noninvariant fermion condensate \cite{peccei_constraints_1977,gasser_quark_1982, brodsky_essence_2010, fukuyama_searching_2012,sikivie_invisible_2020}  our phenomenological model allows to include its effects into the spin dynamics. More, it shows that such a static pseudoscalar could influence the spin motion to some degree.  It is the hypothesis that must be tested and rejected or accepted based on experimental data. With the additional parameter $\beta\ne 0$, there are at least two options available. 
\\[1ex]

\noindent\textbf{Case A:} This option assumes that the effective EDM $d^{exp}$ is not zero
\begin{equation}\label{eq81}
	d^{exp} = d^{th}-\frac{a^{th} e }{2 m}\beta_0 \ne 0\,,
\end{equation}
where $\beta_0$ might not have a direct connection to $d$. Hence, both $d$ and $\beta_0$ are to be found from experiments and tested versus theoretical predictions. This option overlaps with the conventional case, since it includes the possibility that $\beta_0=0$.  Depending on the signs of both $a$ and $\beta_0$, the pseudoscalar correction might reduce or enhance the intrinsic electric moment, similar to the atomic case \cite{ginges_violations_2004}. Since the contributions of intrinsic EDM and the pseudoscalar correction into the linear response to electric fields are viewed as independent in this option, we can estimate the upper bound on $\beta_0$ from the recent EDM test \cite{andreev_improved_2018}. The estimate \cite{porshnev_electron_2020} shows that 
\begin{equation}
	\abs{\frac{a e}{2m}\beta_0} < 10^{-29} e\cdot cm \qquad \to\qquad \beta_0 < 5\times 10^{-16}\,.
\end{equation}
Such a small value of $\beta_0$ makes the pseudoscalar contribution into electron magnetic moments negligible, at least in Case A.  If $\beta_0=0$, then the conventional equation \eqref{eq81} is all that is needed for the phenomenological description of $CP$-violating interactions. The path forward is the same as before - the very sensitive experiments are needed until the breakthrough is achieved in measuring a nonzero effective EDM. Since this case assumes that $\beta_0$ is independent of $d$, our model suggests \cite{baryshevsky_search_2020} that experiments with heavier fermions are favored since the pseudoscalar correction scales down with mass. 
\\[1ex]

\noindent\textbf{Case B:} The second option opens new opportunities and has more predictive power since it is more restrictive. We assume here that the pseudoscalar correction completely offsets the intrinsic EDM
\begin{equation}\label{eq38}
	d^{exp} = d^{th}-\frac{a^{th} e }{2 m}\beta_0 = 0\,.
\end{equation} 
It immediately explains why the EDM has not been detected so far despite the significant improvements in experimental accuracy over several decades. Since we have two terms that contribute into the response to external electric fields, there is a possibility that they cancel each other at rest. The presence of second term that mimics the intrinsic EDM is the common feature in models with $CP$-violating interactions, see \cite{baryshevsky_search_2020}.

Let us pause here to look at the problem from the heuristic point of view, similarly to how the BMT equation was derived in the first place. With only two fundamental constants available (besides $e$ and $m$), the $g$ factor and the electric dipole moment $d$, the BMT equation is the only way to construct a covariant gauge-invariant equation that is linear in both fields and spin.  This equation reduces to the well known spin equation for the nonrelativistic case, since both $g$ and $d$ are taken as nonzero at rest. If our hypothesis has any merit, then there exists an additional parameter $ (g-2) e \beta_0 / 2m\ne 0$. It has the dimension of $d$, transforms as $d$, and it is nonzero at rest. By the nature of $q_0$ which is related to coexistence of charges with opposite signs, as we discussed before, it is conceivable that this new parameter must be added to the permanent EDM $d$ to take it into account. Hence, having three parameters at rest, one can heuristically derive a BMT-like equation; the uncertainty with the heuristic approach would be the sign of term $ae\beta/m$. 

It is more then just the argument based on dimensional grounds. As we discussed before, $q$ is nonzero if a fermion has an additional antifermion component which does not disagree with our current understanding of electrons as surrounded by clouds of virtual pairs. The field theory considers a fermion current as having up to four covariant form-factors \cite{nowakowski_all_2005, eidelman_tau_2016}
\begin{multline}\label{eq34}
\bar{u}({\bf p}_1)\Big[ F_1(k^2)\gamma^\mu
+\frac{i\sigma^{\mu\nu}}{2m}k_\nu F_2(k^2)+ i\epsilon^{\mu\nu\alpha\beta}
\frac{\sigma_{\alpha\beta}}{4m}k_\nu F_3(k^2)\\[1ex]
+ \frac{1}{2m}\left( k^\mu - \frac{k^2}{2m}\gamma^\mu \right)\gamma_5 F_4(k^2) \Big] u({\bf p}_2)\,,
\end{multline}
where $k=p_1-p_2$. The first two form-factors correspond to the charge and magnetic moments, while the third one corresponds to the electric dipole moment. Since the form factors depend on momentum, they will lead to some distributions of electric charge and current in the configuration space. Placed into an external electric field, a permanent dipole without surrounding charges will have to align along the field to minimize its energy. If there exist surrounding charges and currents, re-distributing them will eliminate the dipole energy in the more efficient way.  
 
By using \eqref{eq38},  we can eliminate one parameter, $d$ or $\beta_0$, out of the magnetic coefficients \eqref{eq93}. By eliminating $d$, they become
 \begin{gather}
 \begin{aligned}\label{eq96}
 	&a^{exp} &&= a^{th}\big(1+\beta_0^2 \big)\,,\\[1ex]
 	&a^{*exp}&&=a^{th}\big(1+\frac{3}{4}\beta_0^2 \big)\,.
 \end{aligned}
 \end{gather}
If the pseudoscalar correction is not taken into account, then the absolute value of measured magnetic anomaly will exceed the theoretical value
\begin{equation}\label{eq97}
	\abs{ a^{exp}}   > \abs{ a^{th}}\,.
\end{equation}
Remarkably, the inequality holds independently of signs of magnetic anomaly $a^{th}$, the static pseudoscalar $q_0$, and dipole moment $d^{th}$. Clearly, we assume that both experiment and theoretical evaluation achieved a level where the pseudoscalar correction becomes the primary source of difference $(a^{exp}- a^{th})$. Given a sufficient level of accuracy, this prediction should be expected for other fermions. It is a quite restrictive prediction which is the signature of case B. It also holds for the ultrarelativistic case, if \eqref{eq38} remains valid. 

The latest comparison between experimental and theoretical values for the electron magnetic anomaly\cite{aoyama_theory_2019} shows that
\begin{equation}\label{eq98}
	(a_e)^{exp} - (a_e)^{th} = -8.8 \times 10^{-13}\,.
\end{equation}
which contradicts to the prediction \eqref{eq97}. It still does not reject case B, since the discrepancy \eqref{eq98} might be too high for the electron pseudoscalar correction to play any role. The estimate of electron $\beta_0$ from case A shows that it is too small to be significant here, even if case A is not strictly speaking applicable here. 

This is where the same mass scaling that we  briefly discussed for case A plays the very different role. In case A, we assumed that $d^{th}$ and $\beta_0$ are independent from each other. Hence, a larger mass leads to smaller value of pseudoscalar correction into $d^{exp}$. In case B instead, a larger mass forces a higher value of $\beta_0$ to offset the intrinsic $d^{th}$. In return, a higher value of $\beta_0$ starts playing a more significant role in magnetic coefficients. Let us look at the magnetic anomaly for muon \cite{aoyama_anomalous_2020}
\begin{equation}\label{eq99}
	(a_\mu)^{exp} - (a_\mu)^{th} = 2.8 \times 10^{-9}\,,
\end{equation}
which corresponds to a $3.7\sigma$ discrepancy. Assuming that the pseudoscalar correction is the primary source of this discrepancy, we estimate that 
\begin{equation}\label{key}
	\beta_{0\mu} < 1.6 \times 10^{-3}\,.
\end{equation}
It allows to predict the muon EDM that we get from \eqref{eq38} as
\begin{equation}
	(d_\mu)^{th} = \frac{(a_\mu)^{th} e }{2 m_\mu}\beta_{0\mu}  <1.7 \times 10^{-19}e\cdot cm
\end{equation}
The current upper bound on muon EDM is $1.8 \times 10^{-19}e\cdot cm$ \cite{zyla_review_2020}. A really interesting case is the magnetic anomaly for tau, since the pseudoscalar correction should be even more prominent for the heaviest lepton. Anyway, if case $B$ adequately describes the experimental situation, precise measurements of magnetic anomaly might be an easier way to detect  EDMs which could be completely or partially screened. 

These preliminary estimates have been made for the nonrelativistic case. The screening of EDM might be weakening for relativistic velocities, similar to \cite{commins_electric_2007, senkov_schiff_2008}. This preliminary comment requires an in-depth investigation which we plan to conduct in a future publication. The quantum corrections that we estimated in this study within the second order of WKB approximation are expected to play a role.

A few final comments. The developed phenomenological model allows to potentially explain why the EDM is so difficult to measure and the discrepancy in the muon magnetic anomaly in one framework. Since we suspect the screening of intrinsic $d$, its value can be much higher than the current upper bounds evaluated by means of conventional phenomenological models. Consequently, some of numerous theoretical models of $CP$-violating interactions that were rejected by interpreting the measured upper bounds of EDMs within conventional phenomenological models could come back to life.


\section*{References}

\bibliographystyle{unsrt}

\end{document}